\documentstyle[epsfig]{mn}

\newif\ifAMStwofonts
\AMStwofontstrue

\def\be{\begin{equation}}
\def\ee{\end{equation}}

\def\en{{\cal N}}

\def\etal{{\it et al.~}}

\def\gs{\mathrel{\raise1.16pt\hbox{$>$}\kern-7.0pt
\lower3.06pt\hbox{{$\scriptstyle \sim$}}}}
\def\ls{\mathrel{\raise1.16pt\hbox{$<$}\kern-7.0pt
\lower3.06pt\hbox{{$\scriptstyle \sim$}}}}
\def\gtsima{$\; \buildrel > \over \sim \;$}
\def\ltsima{$\; \buildrel < \over \sim \;$}
\def\prosima{$\; \buildrel \propto \over \sim \;$}
\def\gsim{\lower.5ex\hbox{\gtsima}}
\def\lsim{\lower.5ex\hbox{\ltsima}}
\def\simgt{\lower.5ex\hbox{\gtsima}}
\def\simlt{\lower.5ex\hbox{\ltsima}}
\def\simpr{\lower.5ex\hbox{\prosima}}

\def\pp{\noindent\parshape 2 0truecm 17truecm 2truecm 15truecm}
\def\rf#1;#2;#3;#4 {\par\pp#1, #2, #3, #4. \par}

\def\pr{\ref@jnl{Phys.Rev}}     

\def\ie{{\frenchspacing\it i.e. }}
\def\eg{{\frenchspacing\it e.g. }}

\def\href#1;#2 {{\bf #1} : {\em #2}}

\def\beq#1{\begin{equation}\label{#1}}
\def\eeq{\end{equation}}
\def\beqa#1{\begin{eqnarray}\label{#1}}
\def\eeqa{\end{eqnarray}}

\def\H2p{H$_2^+$ }

\def\mH2p{H_2^+}

\ifoldfss
  \ifCUPmtlplainloaded \else
    \NewTextAlphabet{textbfit} {cmbxti10} {}
    \NewTextAlphabet{textbfss} {cmssbx10} {}
    \NewMathAlphabet{mathbfit} {cmbxti10} {} 
    \NewMathAlphabet{mathbfss} {cmssbx10} {} 
  \fi
  \ifAMStwofonts
    \ifCUPmtlplainloaded \else
      \NewSymbolFont{upmath} {eurm10}
      \NewSymbolFont{AMSa} {msam10}
      \NewMathSymbol{\upi}     {0}{upmath}{19}
      \NewMathSymbol{\umu}     {0}{upmath}{16}
      \NewMathSymbol{\upartial}{0}{upmath}{40}
      \NewMathSymbol{\leqslant}{3}{AMSa}{36}
      \NewMathSymbol{\geqslant}{3}{AMSa}{3E}

       \let\le=\leqslant
       \let\ge=\geqslant
    \fi
  \fi
\fi 

\ifnfssone
  \newmathalphabet{\mathit}
  \addtoversion{normal}{\mathit}{cmr}{m}{it}
  \addtoversion{bold}{\mathit}{cmr}{bx}{it}
  \newmathalphabet{\mathbfit} 
  \addtoversion{normal}{\mathbfit}{cmr}{bx}{it}
  \addtoversion{bold}{\mathbfit}{cmr}{bx}{it}
  \newmathalphabet{\mathbfss} 
  \addtoversion{normal}{\mathbfss}{cmss}{bx}{n}
  \addtoversion{bold}{\mathbfss}{cmss}{bx}{n}
  \ifAMStwofonts
    \ifCUPmtlplainloaded \else
      \UseAMStwoboldmath
      \makeatletter
      \new@mathgroup\upmath@group
      \define@mathgroup\mv@normal\upmath@group{eur}{m}{n}
      \define@mathgroup\mv@bold\upmath@group{eur}{b}{n}
      \edef\UPM{\hexnumber\upmath@group}
      \new@mathgroup\amsa@group
      \define@mathgroup\mv@normal\amsa@group{msa}{m}{n}
      \define@mathgroup\mv@bold\amsa@group{msa}{m}{n}
      \edef\AMSa{\hexnumber\amsa@group}
      \makeatother
      \mathchardef\upi="0\UPM19
      \mathchardef\umu="0\UPM16
      \mathchardef\upartial="0\UPM40
      \mathchardef\leqslant="3\AMSa36
      \mathchardef\geqslant="3\AMSa3E

       \let\le=\leqslant
       \let\ge=\geqslant
    \fi
  \fi
\fi 

\ifnfsstwo
  \DeclareMathAlphabet{\mathbfit}{OT1}{cmr}{bx}{it}
  \SetMathAlphabet\mathbfit{bold}{OT1}{cmr}{bx}{it}
  \DeclareMathAlphabet{\mathbfss}{OT1}{cmss}{bx}{n}
  \SetMathAlphabet\mathbfss{bold}{OT1}{cmss}{bx}{n}
  \ifAMStwofonts
    \ifCUPmtlplainloaded \else
      \DeclareSymbolFont{UPM}{U}{eur}{m}{n}
      \SetSymbolFont{UPM}{bold}{U}{eur}{b}{n}
      \DeclareSymbolFont{AMSa}{U}{msa}{m}{n}
      \DeclareMathSymbol{\upi}{0}{UPM}{"19}
      \DeclareMathSymbol{\umu}{0}{UPM}{"16}
      \DeclareMathSymbol{\upartial}{0}{UPM}{"40}
      \DeclareMathSymbol{\leqslant}{3}{AMSa}{"36}
      \DeclareMathSymbol{\geqslant}{3}{AMSa}{"3E}

       \let\le=\leqslant
       \let\ge=\geqslant
    \fi
  \fi
\fi 

\ifCUPmtlplainloaded \else
  \ifAMStwofonts \else 
    \def\upi{\pi}
    \def\umu{\mu}
    \def\upartial{\partial}
  \fi
\fi

\title[Reionization around the First Stars]{Cosmological Reionization 
Around the First Stars:  Monte Carlo Radiative Transfer}

\author[B. Ciardi, A. Ferrara, S. Marri \& G. Raimondo]{
B. Ciardi$^1$, A. Ferrara$^2$, S. Marri$^3$ and  G. Raimondo$^4$\\  
$^1$ Universit\'a di Firenze, Largo Enrico Fermi 5, 50125 Firenze, Italy \\ 
$^2$ Osservatorio Astrofisico di Arcetri, Largo Enrico Fermi 5, 50125 Firenze, Italy \\  
$^3$ Max-Planck-Institut f\"ur Astrophysik, Karl-Schwarzschild-Stra\ss e 1, 85748 Garching, Germany\\ 
$^4$ Osservatorio Astronomico di Collurania, Via Maggini, 64100 Teramo, Italy\\}

\date{November 2000}

\pagerange{\pageref{firstpage}--\pageref{lastpage}}
\pubyear{2000}
\begin{document}

\maketitle
\label{firstpage}
\begin{abstract}
We study the evolution of ionization fronts around the first
proto-galaxies by using high resolution numerical cosmological 
($\Lambda$+CDM model) simulations and Monte Carlo radiative transfer 
methods.  We present the numerical scheme in detail and show the results of 
test runs from which we conclude that the scheme is both fast and accurate.
As an example of interesting cosmological application, we study  the 
reionization produced by a stellar source of total mass $M = 2\times 10^8 M_\odot$ 
turning on at $z\approx 12$, located at a node of the cosmic web.
The study  includes a Spectral Energy Distribution of a zero-metallicity stellar 
population,
and two Initial Mass Functions (Salpeter/Larson).
The expansion of the I-front is followed as it
breaks out from the galaxy and it is channeled by the 
filaments into the voids, assuming, in a 2D representation, 
a characteristic butterfly shape. The ionization evolution is very
well tracked by our scheme, as realized by the correct treatment of 
the channeling and shadowing effects due to overdensities. 
We confirm previous claims that
both the shape of the IMF and the ionizing power metallicity dependence are 
important to correctly determine the reionization of the universe.
\end{abstract}
\begin{keywords}
galaxies: formation - intergalactic medium - cosmology: theory
\end{keywords}

\section{Introduction}

An increasing number of works has been recently dedicated to the study 
of the reionization of the universe (Gnedin \& Ostriker 1997; 
Haiman \& Loeb 1998; Valageas \& Silk 1999; Ciardi \etal 2000, CFGJ; 
Miralda-Escud\'e, Haehnelt \& Rees 2000; Chiu \& Ostriker 2000; 
Bruscoli \etal 2000; Gnedin 2000; Benson \etal 2000) which use  both
analytical and numerical approaches. An important refinement
has been introduced by the proper treatment of a number of feedback effects
ranging from the mechanical energy injection to the H$_2$ photodissociating 
radiation produced by massive stars (CFGJ).
However, probably the major ingredient still lacking for a physically complete
description of the reionization process is the correct treatment of the
transfer of ionizing photons from their production site into 
the intergalactic medium (IGM).
Attempts based on different approximated techniques have 
been proposed 
(Razoumov \& Scott 1999; Abel, Norman \& Madau 1999; Norman, Paschos \& Abel 1998; Gnedin 2000; Umemura,
Nakamoto \& Susa 1999), 
which sometimes are not readily implemented in cosmological simulations.
Therefore, it is crucial to develop exact and fast methods that can 
eventually yield a better treatment of the propagation of ionization
fronts in the early universe. 
The first three papers are based on the ray-tracing
method, whereas Gnedin (2000) uses the so called local optical depth approximation.
Finally Umemura, Nakamoto \& Susa (1999) implemented a time-independent ray-tracing method.
Monte Carlo (MC) methods have been widely used in several physical/astrophysical
areas to tackle radiative transfer problems (for a reference book see 
Cashwell \& Everett 1959) and they have been shown to result in fast and accurate
schemes. Here we build up on previous experience of our group (Bianchi,
Ferrara \& Giovanardi
1996; Ferrara \etal 1996; Ferrara \etal 1999; Bianchi \etal 2000)
in dealing with MC  problems to present a case study of
cosmological H$_{\rm II}$ regions produced by the first stellar sources.     
The study is intended as a test of the applicability of the adopted
techniques in conjunction with cosmological simulations in terms of
convergency, accuracy, and speed of the scheme. 

As an example of interesting cosmological application, we study  the
reionization produced by a stellar source of total mass $M = 2\times 10^8 M_\odot$
turning on at $z\approx 12$, located at a node of the cosmic web.
The study  includes a Spectral Energy Distribution of a zero-metallicity stellar population,
and two Initial Mass Functions (Salpeter/Larson);
the IGM spatial density distribution is 
deduced from high resolution cosmological simulations described below.

In a forthcoming paper we will then improve the
results of CFGJ by exploiting the strength of the MC method
to release the approximations relative to the radiative transfer made in
that paper. 

\section{Numerical Simulations}

We have studied the evolution of small scale cosmic structure in a
standard $\Lambda$+CDM model with $\Omega_\Lambda =0.7$ and $\Omega_{m} =0.3$.
The baryon contribution to $\Omega_{m}$ is $\Omega_{b}=0.04$, the adopted
value of the Hubble constant is $H_{0}=70$~km~sec$^{-1}$~Mpc$^{-1}$.

The initial conditions are produced with the
COSMICS \footnote{Bertschinger, http://arcturus.mit.edu/cosmics/} code
adopting a normalization $\sigma_{8}=1.14$ for
the power spectrum. The initial particle distribution (at redshift $z=103.2$)
is made out of a $256^3$ grid in a box of $4$~Mpc (comoving),
thus giving a mass per particle of about $1.5 \times 10^{5} M_{\odot}$.
We adopt vacuum boundary conditions; this requires the initial box
to be reshaped.  Assuming that tidal forces from large scale fluctuations
are negligible for this problem, we
cut a sphere of radius $2$~Mpc from the initial cube and follow the evolution
of the system in isolation. In practice, we study the evolution of a central
sphere of radius about $1.2$~Mpc in full mass resolution with a coarse
resampling of the most external particles. In the hi-resolution
(central) region the number of particles is 2313847,  whereas the rebinning
left us with 10008 particles of variable mass
(increasing with distance from the center) in the external $1.2-2.0$~Mpc shell.

We evolve this initial matter distribution with GADGET (Springel, Yoshida \&
White 2000), a parallel N-body tree-code with an SPH scheme to describe
hydrodynamical processes.
SPH particles are placed on top of hi-resolution dark matter particles
with the same velocity and a low initial temperature
(about $150$~K, the temperature of intergalactic medium
 as determined by adiabatic expansion of the universe after decoupling);
the mass of each fluid particle is rescaled according
to the baryon and matter fraction quoted above.
The (spline) softening length is $1$~kpc for the gas particles, and
$3$~kpc for hi-resolution dark matter particles, respectively.
For the problem at study here, the simulation is stopped at $z \approx 5$,
thus avoiding the very strong nonlinear phase in which  our assumption of
zero tidal effects from large scale structure would break down.

The gas-dynamics evolution is purely adiabatic in order to reasonably limit the 
integration time. Thus, we are not able to follow the details of
cooling processes that ultimately lead to the formation of very dense and cold
star forming regions. However, the density field of the mildy overdense IGM 
should be reasonably well described by this assumption. Our mass resolution 
does not also allow a full resolution of the Jeans mass, and therefore the 
presence of sub-grid structure that we cannot track is expected. 
This is a problem common to all presently available simulations,
implying that the treatment of IGM clumping is only approximate, as recently pointed 
out by Haiman, Abel \& Madau (2000).

Simulations were performed on the Cray-T3E at Rechenzentrum-Garching, the
Joint Computing Center of the Max-Planck-Gesellschaft and Max-Planck-Institut
fuer Plasmaphysik.

Finally, from the SPH particle temperature and density distribution we
reconstruct a central cubic region of $128^3$ cells, the grid values of
any field being evaluated from particle values with an SPH-like interpolation.
For the reionization study presented   below, we have concentrated, as
an example,  on a source located at redshift $z \approx 12$ immersed in the
density field calculated from the simulations described above. The
location and mass of the dark matter halos in the simulation have been
determined by means of a friend-of-friend algorithm. We have then selected a halo suitably
located in the node of a cosmic filamentary structure and associated a
luminous source to it. The total halo mass is $M=2 \times 10^8 M_\odot$; the
corresponding stellar mass has been determined following the same
prescriptions of CFGJ, \ie
\begin{equation}
M_{\star}=(\Omega_b/\Omega_m) M f_b f_{\star},
\label{mstar}
\end{equation}
where
$f_b=0.08$ is the fraction of virialized baryons able to cool and
become available to form stars, and $f_{\star}=0.15$ is the star formation
efficiency. 
Although with our cell size of 15 kpc  we are marginally able to
resolve the source (comoving) virial radius (about 20 kpc),
we do not attempt to treat the  the escape of ionizing photons
(Dove, Shull \& Ferrara 2000; Wood \& Loeb 2000)
from the galaxy interstellar medium. Instead we take a fixed value
for this quantity equal to $f_{esc}=0.2$, the upper limit obtained by
the above studies, by which we multiply the source 
ionizing photon flux. 

\section{PopIII Star Spectrum and IMF}

\begin{figure}
\vskip -2truecm
\psfig{figure=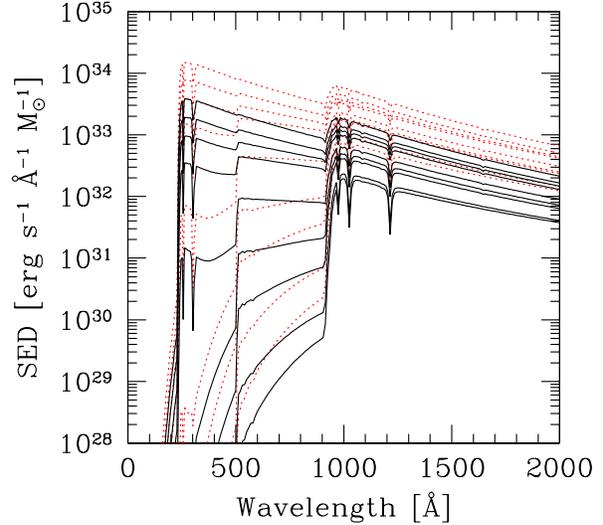,height=10cm}
\caption{\label{fig1}\footnotesize{Evolution of the SED for a Simple
Stellar Population of metal free stars for two different IMFs: Salpeter
(solid lines) and Larson with $M_c=5 M_\odot$ (dotted). The evolutionary
times are, from the top to the bottom, $5, 7, 10, 15, 30, 50, 70, 100,
150$~Myr.}}
\end{figure}

The SEDs of metal free star systems used      in
this paper were calculated implementing the synthetic evolutionary
code developed for a Simple Stellar Population (Brocato \etal 1999,
Brocato \etal 2000). For a detailed description of the computational
procedure we refer to those papers; here we just outline that the results
of this code have been largely tested on young LMC clusters and 
old galactic globular clusters showing a very good agreement.  The
present synthesis models rely on the set of homogeneous evolutionary
computations for stars with a cosmological amount of heavy elements 
and helium (Z=10$^{-10}$ and Y=0.23) presented by
Cassisi \& Castellani (1993) and Cassisi, Castellani \& Tornamb\`e (1996) and properly
extended to higher masses in order to cover a wide range of ages.  
All the major evolutionary phases are taken into account, both the H and the He
burning phase for stars with original masses in the range 0.7$-$40
M$_{\odot}$, for which the He burning phase is followed up to the
Carbon ignition or, alternatively, to the onset of the thermal
pulses. Models atmospheres are by Kurucz (1993) for metallicity
Z=2$\times 10^{-7}$. The available temperatures range from 3700~K to
50000~K, so the spectra at higher temperatures and gravity are simply
extrapolated in order to keep the homogeneity of the grid. The
wavelength range is from 91~\AA~  to 160~$\mu$m and the spectral
resolution is typically 10-20~\AA~  in the UV to optical bands. We
consider an instantaneous burst of star formation assuming that 
masses range from 1~$M_\odot$, as recently suggested for Pop~III stars
by Nakamura \& Umemura (1999a, 1999b), to 40~$M_\odot$ and two cases for the
IMF: $i)$ a standard Salpeter law,
and $ii)$ a Salpeter function at the 
upper mass end which falls off exponentially below a characteristic 
stellar mass $M_c$ (Larson 1998). The last case 
is taken into account since both theory of thermo-dynamical condition of primordial gas and the
absence of observations of solar mass, metal free stars seem to suggest
at birth a paucity of low mass stars.
Moreover, the mass interval used in this paper properly  
describes population models of
age down to 5~Myr, the lifetime of a 40 M$_\odot$ star;
therefore, the burst models of this age are actually populated by
1-40~M$_\odot$ stars. The resulting grid of SEDs             
were calculated at intervals of 1, 10 and 50~Myr between
1 and 20~Myr, 20 and 100~Myr, and 100 and 150~Myr, respectively.
Other calculations of zero-metallicity star SEDs are present in the
literature (see e.g. Cojazzi \etal 2000; Tumlinson \& Shull 2000), 
but they do not show the detailed SED time evolution. 

\begin{figure}
\vskip -2truecm
\psfig{figure=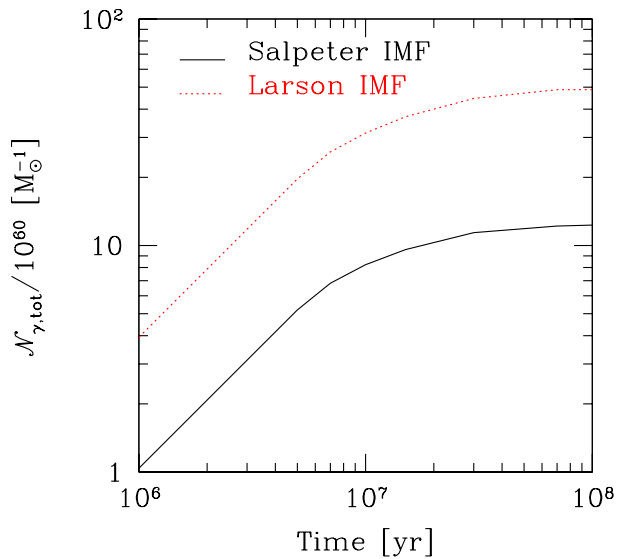,height=10cm}
\caption{\label{fig2}\footnotesize{Time evolution of the total number of
ionizing photons (in units of $10^{60}$) emitted by a solar mass of metal
free stars with a Salpeter (solid line) IMF and a Larson (dotted) IMF with $M_c=5M_\odot$.}}
\end{figure}

In Fig. \ref{fig1} we show the adopted SEDs for the Salpeter IMF and the
Larson IMF with $M_c=5 M_\odot$. The two sets of curves do not differ 
sensibly in their spectral shape evolution but the Larson IMF has
about 4 times larger specific photon flux at the Lyman continuum.
In Fig.~\ref{fig2} we show the time evolution of the total number of
ionizing photons (in units of $10^{60}$) emitted by a solar mass of metal
free stars with the two IMFs. After about $10^7$~yr both curves 
flatten as a consequence of the decreasing number of surviving  
massive stars. 
It is important to note that, during the first 5 Myr, a zero-metallicity 
stellar cluster with a Salpeter IMF has an ionizing photon rate about 25 
times higher than the analogous one for a 1\% solar metallicity used by CFGJ.
As we will see later, this has important effects on the reionization
process.

\section{Monte Carlo Radiative Transfer}

The application of MC schemes to radiative transfer problems requires
that the radiation field is discretized in a representative number of 
photon packets, $\en_{p}$. The processes involved
(\eg packet emission and absorption) are then treated statistically by randomly
sampling the appropriate distribution function. 
Let us consider a source of bolometric luminosity
$L$ and lifetime $t_s$ ($L$ can be a function of time as our method easily 
allows to treat short-lived and variable sources). 
Packets have the same energy, ${\cal E}_p$,  but contain a different number of 
monochromatic photons, $N_\gamma$, of frequency $\nu$  so that for
the $j$-th packet it is ${\cal E}_p = N_\gamma(j) h \nu(j)$; their rate is
$\dot {\cal N}_p=L/{\cal E}_p$ and  the simulation time when the $j$-th
photon packet is emitted is then $t= j dt$ ($j=1,..,{\cal N}_p$),
where $dt = 
\dot {\cal N}_p^{ -1}$. 
The number of photons in each packet emitted by the source during $t_s$        is:
\be
\label{fotnumb}
N_\gamma(j) = {E_s\over {\cal N}_p h \nu(j)} ={L t_s\over {\cal N}_p h \nu(j)}.
\ee
The packet frequency, $\nu$, is obtained by sampling the source 
SED, $S(\nu)$: if  $R_\nu$ is a
random number, $\nu$ is given implicitly by:
\be
R_\nu=\int_{\nu_H}^\nu d\nu' S(\nu') \; \left [ \int_{\nu_H}^{\nu_{max}}
d\nu' S(\nu')  \right ]^{-1},
\label{nu}
\ee
where $\nu_H$ is the threshold for hydrogen ionization.

We adopt spherical coordinates ($r, \theta, \phi$) with origin at the source
location ($x_s, y_s, z_s$) in the box,  such that the cartesian 
coordinates are:  
\begin{eqnarray}
\label{frame}
x & = & r \, {\rm sin}\theta \, {\rm cos}\phi  ~~~0 \le \phi \le 2\pi, \nonumber \\
y & = & r \, {\rm sin}\theta \, {\rm sin}\phi  ~~~0 \le \theta \le {\pi}, \\
z & = & r \, {\rm cos}\theta. \nonumber
\end{eqnarray}

Assuming that the emission is isotropic and spherically 
symmetric, photon packets will propagate  along the direction:
\be
\label{dir}
\hat{\bf s} \equiv [\hat r, ~{\rm arccos}(1-2R_\theta)\hat \theta, ~2 \pi R_\phi\hat \phi],
\ee
where $R_\theta$ and $R_\phi$ are random numbers. Note that the above choice
automatically ensures a constant surface density of photon packets, thus
preventing flux anisotropies at the poles due to angle sampling;  
this can be seen as follows. As we are sampling $R_\theta$ from a uniform distribution, 
the fraction of photon packets emitted in the angular range $\theta_1 < \theta < \theta_2$ 
is equal to 
\be   
\label{sphotden}
d{\cal N}_p(\theta_1,\theta_2) = (R_{\theta_2}-R_{\theta_1}), 
\ee 
where $R_{\theta_i}=(1/2)(1-\cos{\theta_i})$ from eq. \ref{dir}.
As the area fraction of the sphere belt delimited by $\theta_1$ and $\theta_2$ is
\be   
\label{sphotden}
dA(\theta_1,\theta_2) = {1\over 2} (\cos{\theta_1}-\cos{\theta_2}) 
\ee 
one sees that the surface density $d{\cal N}_p/dA$ of photon packets is constant 
and independent  of angles. 

Once emitted, a packet of frequency $\nu$ 
will travel for a finite path length of optical depth:
\be
\label{eqtau}
\tau_i(\nu)=\sigma(\nu) \sum_{k=1}^i (N_{\rm HI})_k = \sigma(\nu) \sum_{k=1}^i (n_{\rm HI})_k f \Delta x, 
\ee
before being absorbed in the $i$-th grid cell. Here $\sigma(\nu)$ is the photoionization 
cross section, $(N_{\rm HI})_k= (n_{\rm HI})_k f\Delta x$ is the IGM neutral hydrogen column 
density through the $k$-th cell of size $\Delta x$; the cell with $k=0$ contains the source
location and it is not included in the opacity count, as its contribution is already included in $f_{esc}$. 
For simplicity, the results 
presented here only include H opacity; inclusion of He, molecules (as H$_2$ and HD)
and heavy elements, if present, is straightforward, although computationally
expensive for molecules as one has to properly compute both the line widths
and their Doppler shifts. 
The factor $f$ accounts for the fact that the path through a cell is in the
range $0 < \ell \le \sqrt{3}\Delta x$ depending on its inclination. 
\begin{figure}
\vskip -2.truecm
\psfig{figure=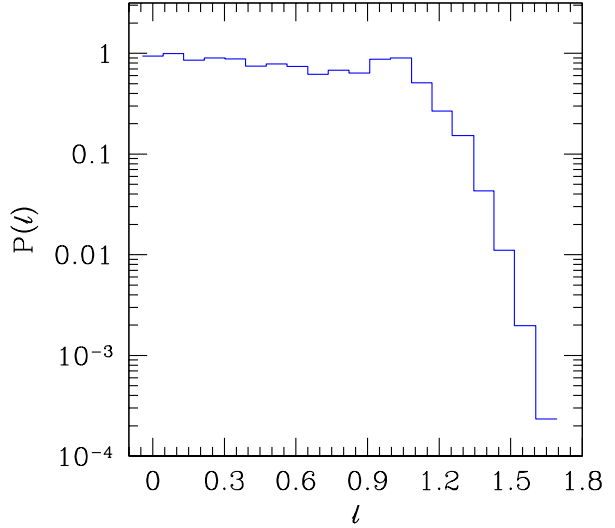,height=10cm}
\caption{\label{fig3}\footnotesize{Differential probability distribution function of lengths, $\ell$
for randomly oriented paths entering from a   face of a cubic box of unit size.
The integral of the curve is normalized to unity.}}
\end{figure}
To limit the computational cost, we do not attempt to calculate the actual path and
we treat this effect statistically as follows. 
Given a cubic box of unit size $\Delta x=1$, via an independent Monte Carlo
procedure, we randomly pick up on a face
the coordinates of an entering ray and the direction of its propagation. We
then derive the differential probability distribution function,  $P(\ell)$, of the path
lengths (Fig.~\ref{fig3}) 
and define $f = 0.56$ as the median value of such distribution.     

The value of $i$ in eq. \ref{eqtau} is defined as the minimum one for
which the following condition is satisfied:
\be
\tau_i(\nu) \ge \tau=-{\rm ln}(1-R_\tau),
\ee
with $R_\tau$ a random number. The previous relation is obtained by sampling the photon absorption 
probability distribution. 

In each cell $k=1,..,i$ along the path from the source to the absorption site
of the $j$-th packet, we update the value of the hydrogen ionization fraction; 
$x=n_{{\rm H}^+}/n_{\rm H}$, $n_{{\rm H}^+}$ and $n_{\rm H}$ are   the ionized and total hydrogen number
density, respectively. This is done by advancing the time-dependent ionization 
equation in discretized form.  At time $t=t^{n+1}$ it is: 
\be
\label{ioneq}
x_k^{n+1}=x_k^{n}+\Delta t [\gamma_c(x_k^{n}) - \gamma_r(x_k^{n})] + 
{N_{\gamma}(j)\over (n_{\rm H})_k  (\Delta x)^3} \delta_{ki};
\ee
$\gamma_c$ and $\gamma_r$ are the collisional ionization and recombination rates,
respectively (Cen 1992); they are evaluated at the temperature of the gas
in the cell. The initial temperature is increased to  $T=10^4$~K  after absorption
of a photon packet and set equal to $T =2.734(1+z)$, \ie the CMB temperature, 
after the cell recombines (see below). 
This is a reasonable approximation,
although $T$ could be smaller at higher redshift (Ciardi, Ferrara \& Abel 2000).
The last term,  describing photoionization       
due to photon packet absorption, is
present only for the cell in which absorption takes place ($k=i, \delta_{ki}=1$);
in the other cells along the path ($k < i, \delta_{ki}=0$) we simply follow   
the gas recombination. 
In case the derived value of $x_k^{n+1} > 1$, we set  $x_k^{n+1} = 1$,
and we use the extra photons to ionize the cell $x_{k+1}$. In practice,
we write $N_{\gamma}(j) = N^{(1)}_{\gamma}(j) + N'_{\gamma}(j)$, where 
the number of photons required to obtain $x_k^{n+1} = 1$ is:
\be
 N^{(1)}_{\gamma}(j)=\{( 1 -x_k^{n})-\Delta t [\gamma_c(x_k^{n}) - \gamma_r(x_k^{n})]\}
(n_{\rm H})_k  (\Delta x)^3. 
\ee
The most delicate point of the ionization fraction evaluation is to find 
an appropriate expression for $\Delta t$. In principle, 
one could  update the ionization state on the entire grid after each photon packet emission.
However, this is computationally too expensive. For this reason we choose
to update the ionization state of cells along the packet path and we deduce $\Delta t$ as follows.

The cell in which the source is located is enclosed by a cubic surface (shell).
If we define the {\it order} of this shell  to be $g=1$, the number of its 
member cells is given by:
\be
N_s(g) = (2g+1)^3 - (2g-1)^3 = 2 (12 g^2 +1) ~~~~~ g\ge 1,
\ee
or $N_s(1) = 26$. The second order shell ($g=2$) is instead made of 98 cells
and so on. As the photon packet directions are isotropically distributed, statistically
a cell in a shell  of order $g$ will be along a path every $[N_s(g)]^{-1}$ packets. 
Therefore the typical time scale between two subsequent ionization fraction updates
is $\Delta t = N_s(g) dt$; this is the value we use in eq.~\ref{ioneq}
above. The order $g$ is calculated as:  
\be
g = {\it Max} [ k_x,  k_y,  k_z],
\ee
where the $k_i$'s indicate the discrete cartesian coordinates of the cell 
in the reference frame eq.~\ref{frame}.
The above approximation sets a lower limit on ${\cal N}_p$, as in order to
be valid it is necessary that the update time $\Delta t$ is shorter than the 
average recombination time, $\langle t_{r} \rangle$, in the box. Hence         
to advance the simulation up to a time $t_s$ the number 
of packets required is:
\be
{\cal N}_p \simgt 24 N^2 {t_s \over \langle t_{r} \rangle}, 
\ee
where $N^3$ is the total numer of cells in the box.
This detailed treatment of recombination is important to accurately evaluate
the optical depth through eq.~\ref{eqtau}. 

We then introduce an ionization vector (size $N^3$) whose elements are initially set to 
zero. The vector element corresponding to a cell for which the ionization fraction
is increased above $x_{th}=0.99$ 
as a result of an absorption event is updated to contain the value:   
\be
j_r = {\it Int}\left[{t_r(x_k, n_k) + t\over dt}\right], 
\ee
where $t_r(x_k, n_k)$ is the local recombination time  and
$t$ is the current simulation time. If further absorption does not take place before
the time $j_r dt$, we allow for recombination emission from the given
cell; if instead another absorption event takes place, we re-update the 
vector element to a
new value of $j_r$. As already mentioned, after recombination, the cell temperature is set at 
the CMB temperature at that redshift. 
 
Following a cell recombination a packet is emitted along a random direction
from the same location. 
The probability that the emitted packet has $h\nu(j) > 13.6$~eV is given by 
the ratio between the number of recombinations to the ground level
and the total one, \ie 50\% for $T=10^4$~K.
The frequency of the re-emitted packet is determined through
eq.~\ref{nu}, where the source SED is substituted by the volume
emissivity of the diffuse radiation, given by the Milne relation:
\begin{eqnarray}
S_d(\nu)& = & \frac{2 h \nu^3}{c^2} \frac{g_i}{g_{i+1}} \left (\frac{h^2}{2 \pi
m_p k T_e} \right )^{3/2}\times \nonumber \\
      &   &  \;\;\;\;\;\;\;\;\;\;\;\;\;\;\;\;\;
      \sigma(\nu) {\rm e}^{-h(\nu-\nu_H)/(k T_e)} n_{H^+} n_e,
\end{eqnarray}
where $g_i$ and $g_{i+1}$ are the hydrogen ground level statistical weights;
$T_e$ and $n_e$ are the electron temperature and number density. 
The propagation and absorption of re-emitted packets is then followed 
in the same manner as for primary ones. The only difference is that
the opacity contribution of the re-emitting cell is now included.  
      
The method is easily generalized to include an arbitrary number of point 
sources once the packet emission rates, and hence their emission sequence  
for the various sources are given.

\begin{figure}
\vskip -2truecm
\psfig{figure=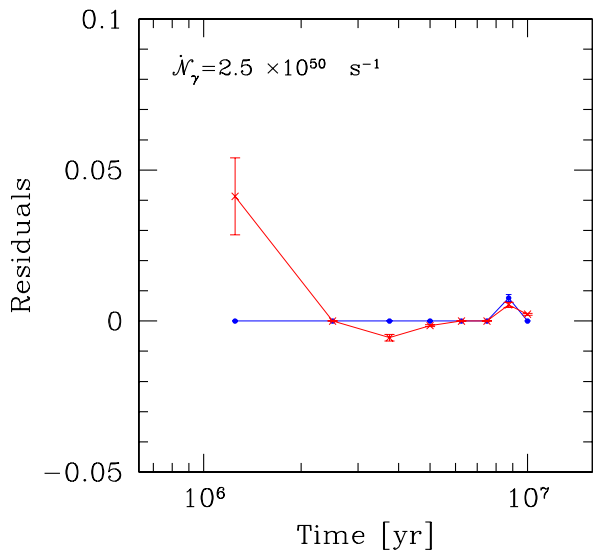,height=10cm}
\caption{\label{fig4}\footnotesize{Residuals of the numerical 
H$_{\rm II}$ region equivalent radius, $R_n$, for the convergency test:
crosses refer to $D^5$, circles to $D^{50}$ (see text for definition of $D$).
The source has ionizing photon rate $\dot{{\en}}_\gamma=2.5 \times 10^{50}$
s$^{-1}$.}}
\end{figure}

\subsection{Testing the Method}

\begin{figure}
\vskip -2truecm
\psfig{figure=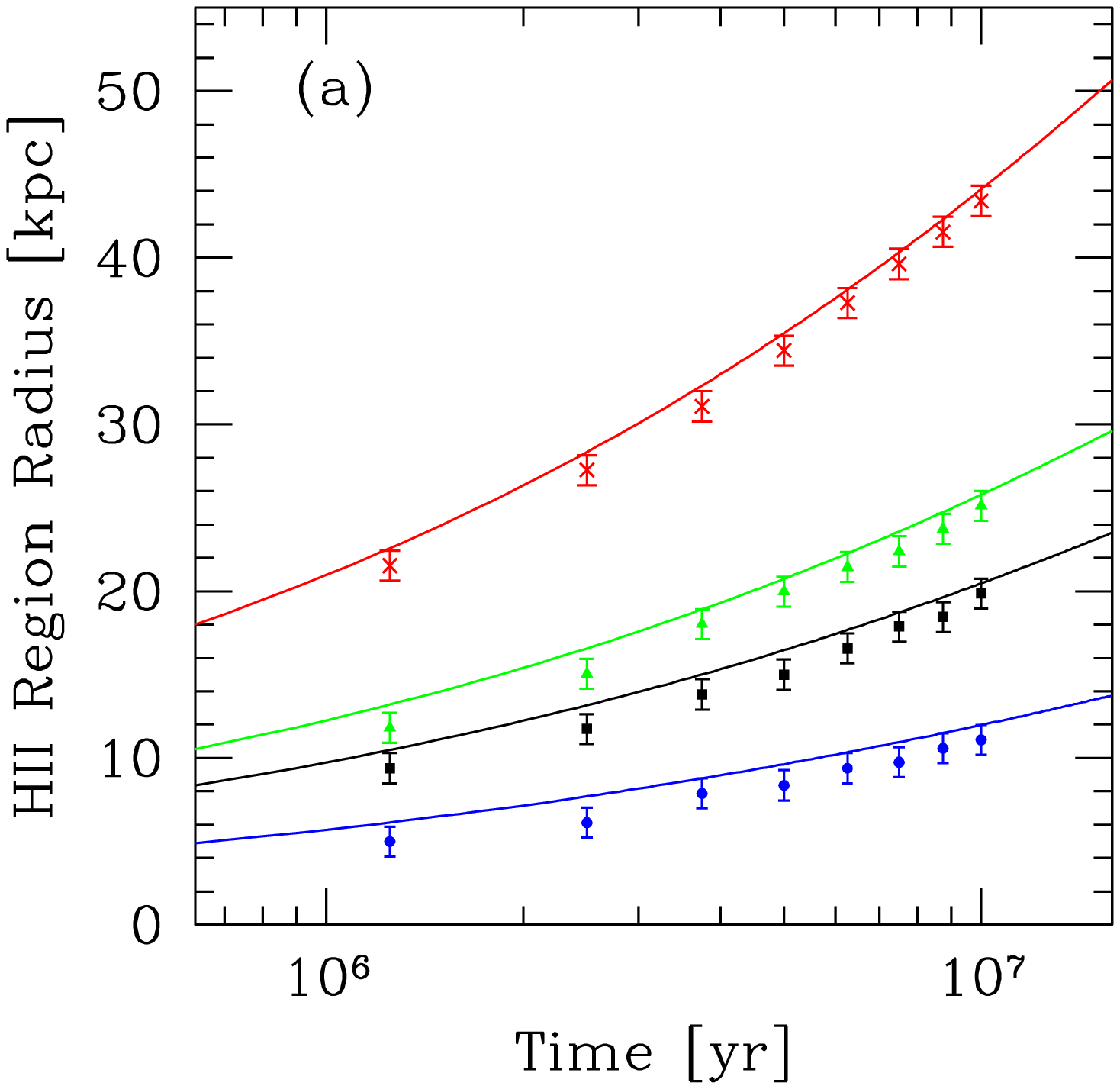,height=10cm}
\vskip -3truecm
\psfig{figure=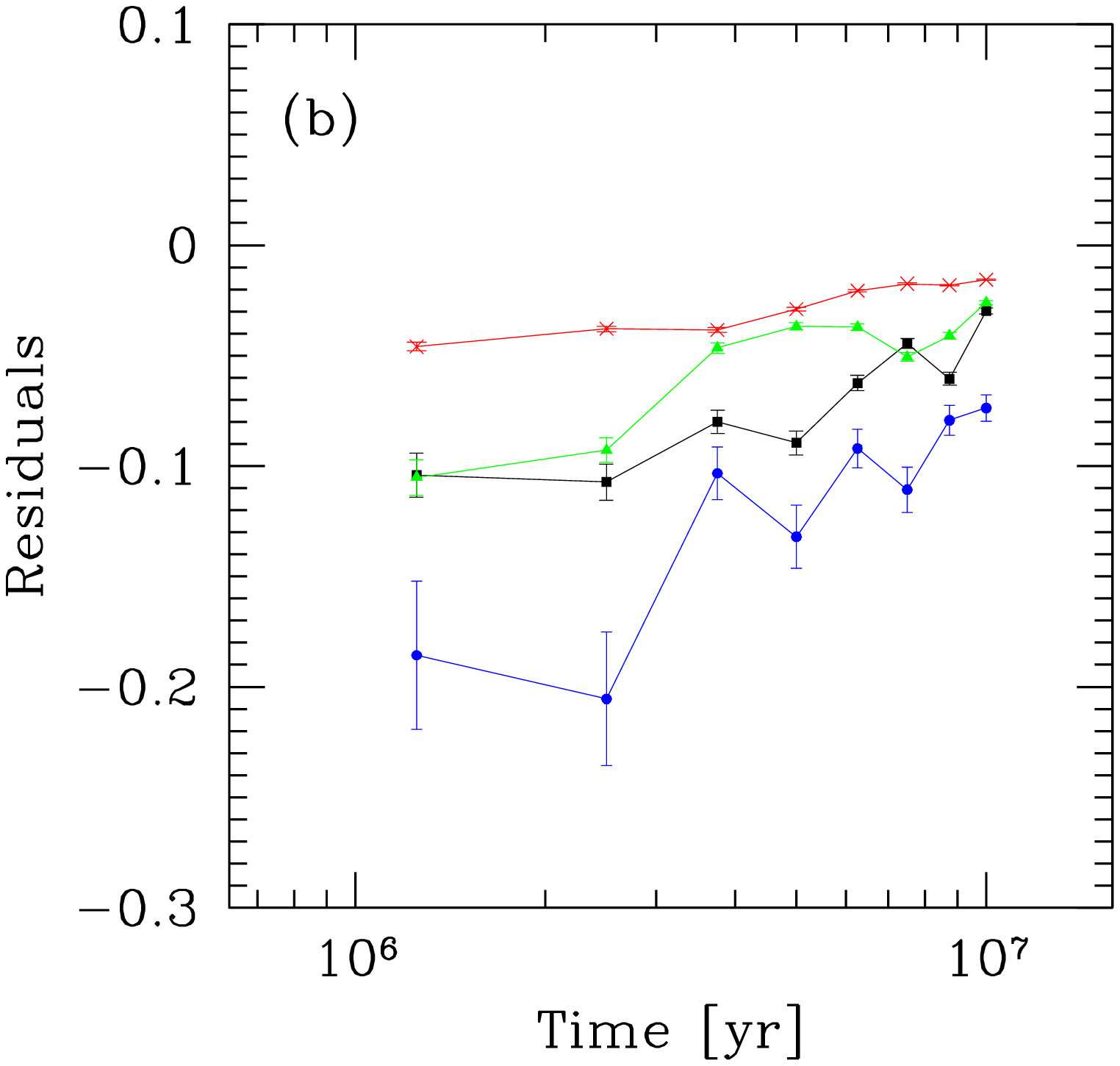,height=10cm}
\caption{\label{fig5}\footnotesize{(a) Time evolution of the equivalent
H$_{\rm II}$ radius produced by sources of different ionizing photon rate
$\dot{{\en}}_\gamma=2.5 {\rm (circles)}, 12.5
{\rm (squares)}, 25 {\rm (triangles)}, 125 {\rm(crosses)}\times 10^{50}$~s$^{-1}$,
compared to the analytical solutions
(solid lines). (b) Residuals of the numerical radii  
calculated with respect to the analytical ones. The notation is the same
as in Fig.~(a).}}
\end{figure}
The method described in the previous Section has been tested against the
analytical solution for the temporal evolution of the radius of an H$_{\rm II}$
region produced by a source with constant ionizing rate and
expanding in an homogeneous medium (Donhaue \& Shull
1987; Shapiro \& Giroux 1987). The H$_{\rm II}$ region equivalent radius, 
$R_{n}$, is numerically derived by first
assuming that a cell $k$ is in the ionized state when $x_k  > x_{th} = 0.99$.
If $N_{ion}$ is the number of ionized cells in the box,
the volume occupied by the ionized region is $V=\Delta x^3 N_{ion}$
and the equivalent radius is then $R_{n}=(3V/4 \pi)^{1/3}$.
In all the tests presented here the gas density is set equal      to 
average IGM one in an Einstein-de Sitter universe with $\Omega_b h^2=0.019$ at $z=11.6$, 
\ie $n_{\rm H} = 4.5 \times 10^{-4}$~cm$^{-3}$; 
the adopted spectrum is monochromatic, with a photon energy
equal  to 13.6~eV.

First, we check for convergency of our scheme. We run three simulations 
with different values of ${\cal N}_p = (5, 50, 500)\times 10^5$.  
The source ionizing photon rate is $\dot{{\en}}_\gamma=
2.5 \times 10^{50}$~s$^{-1}$.
For each run we calculate the residuals with respect to the highest 
resolution run ${\cal N}_p = 500\times 10^5$; they  are defined as:
\be
\label{eqres}
D^y={R_{n}^y-R_{n}^{500} \over R_{n}^{500}},
\ee
where the superscript $y=5,50$ refers to the low- and medium-resolution runs,
respectively. The residuals are shown in Fig.~\ref{fig4} at different evolutionary times 
together with the associated mean quadratic errors; the error on $R_n$ is 
assumed to be equal to $\Delta x$. Satisfactory convergency is achieved  
by the medium-res run; subsequent tests have hence been done using 
${\cal N}_p=50 \times 10^5$. For reference, this run takes only $\simeq 300$~s  
(CPU time) on a SUN ULTRA10/333MHz workstation.

Next, we assess the accuracy of the solution. 
In Fig.~\ref{fig5}a we show the time evolution of $R_{n}$, as produced by
sources of different ionizing photon rates $\dot{{\en}}_\gamma=
(2.5,  12.5, 25, 125) \times 10^{50}$~s$^{-1}$, 
compared to the corresponding analytical radius, $R_a$. 
Fig.~\ref{fig5}b shows the residuals 
(defined analogously  to eq.~\ref{eqres} with $R_n^{500}$ 
substituted by $R_a$)  with respect to the analytical solution.  
In the worst case (very faint source, early evolutionary times) the residual is about 20\%;
This relatively large discrepancy occurs because the size of the ionized region is 
only a few cells. At later times and/or for more luminous sources the 
agreement is remarkably good, \ie within a few percent. 

\begin{figure}
\vskip -2truecm
\psfig{figure=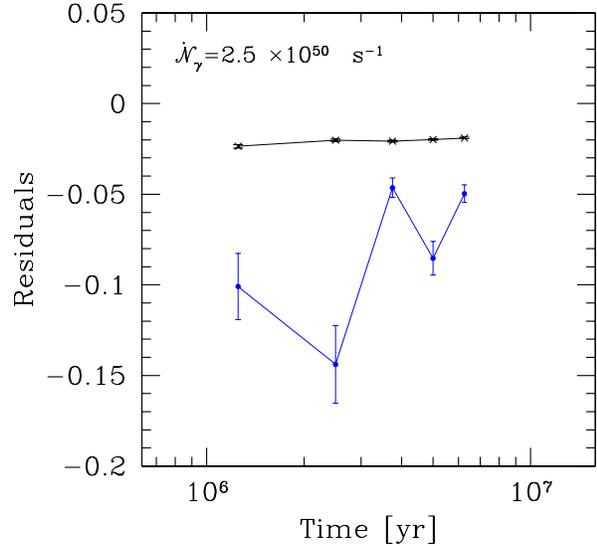,height=10cm}
\caption{\label{fig6}\footnotesize{Residuals
for  runs with
a comoving box size $(1, 0.2) \times N \Delta x$ (circles and crosses, respectively)
with respect to one with $N \Delta x/10$.
The source ionizing photon rate is $\dot{{\en}}_\gamma=
2.5 \times 10^{50}$~s$^{-1}$.}}
\end{figure}

Finally, we check the effects of varying the spatial resolution. To this aim 
we have considered runs with different box sizes, keeping a fixed linear number of  
cells $N=128$. In Fig.~\ref{fig6} we show the residuals for runs with
a box size $ (1, 0.2, 0.1) \times N \Delta x$ with respect to the highest 
spatial resolution run for a source ionizing photon rate   
$\dot{{\en}}_\gamma= 2.5 \times 10^{50}$~s$^{-1}$. 

\begin{figure}
\psfig{figure=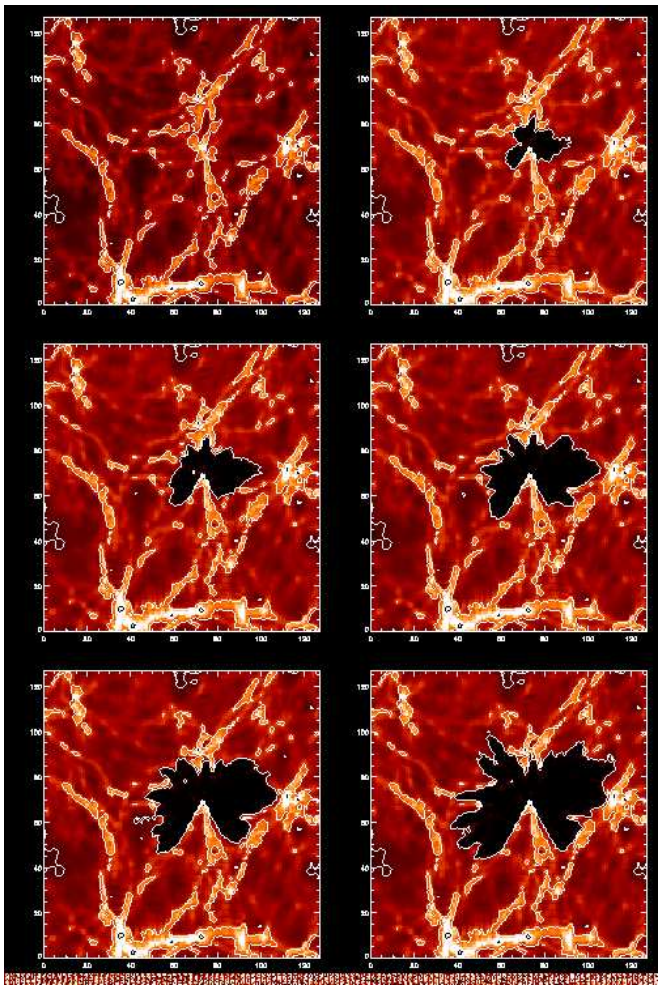,height=13cm}
\caption{\label{fig7}\footnotesize{Slices through the simulation box
containing a metal free stellar source of total mass $M = 2 \times
10^{8} M_\odot$ at $z \approx 12$ in a
$\Lambda$+CDM cosmological density field and a Larson IMF. 
The six panels show the neutral hydrogen number density, superposed to
which (black regions) is the evolution of the ionized ($x > x_{th}=0.99$) 
hydrogen distribution at times 0, 10, 20, 40, 60, and 100 Myr after source turn on. 
The linear (proper) size of each slice is 160 kpc.
}}
\end{figure}
\section{Results and Discussion}

\begin{figure}
\hskip 1.truecm
\psfig{figure=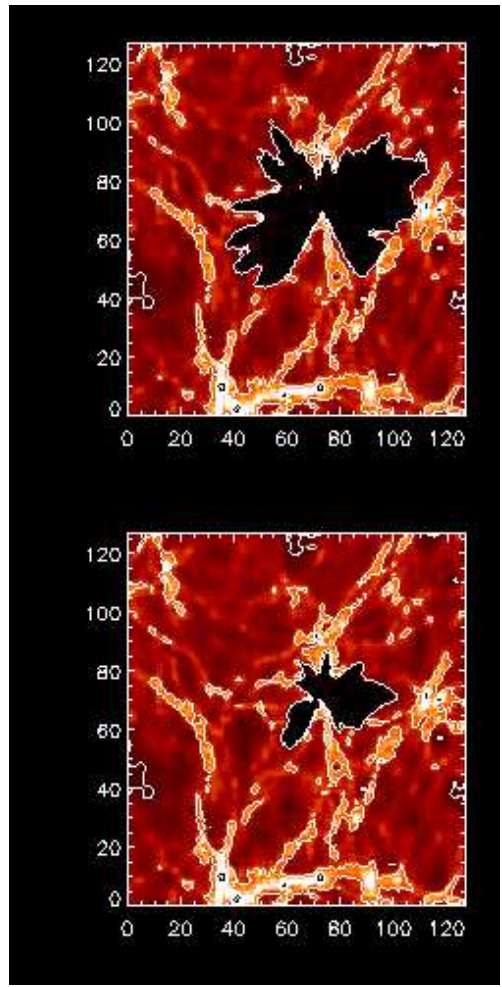,height=13cm}
\caption{\label{fig8}\footnotesize{Comparison between the
final stages ($t=100$~Myr) of the I-front evolution
for a Larson   (upper panel) and a Salpeter (bottom) IMF for the
same source properties as in Fig. \ref{fig7}.}}
\end{figure}
The combined cosmological and radiative transfer simulations described
above allow us to determine the evolution of the ionized region around
the selected zero-metallicity stellar object. Illustrative slices extracted
from our simulation box through the source location are shown in Fig.
\ref{fig7} at different times (10, 20, 40, 60, 100 Myr) 
after the source 
has been turned on and a for Larson IMF, 
together with the initial density field. 
The above time interval corresponds to less than two redshift units
at $z\approx 12$, and it is therefore not 
unreasonable for the purposes of this paper to neglect the IGM density evolution
As seen from Fig. \ref{fig7}, the ionization front (I-front) breaks out 
from the galaxy very rapidly, leaving behind a very clumpy ionization
structure in the immediate surroundings where the IGM is overdense and,
consequently, recombination times are shorter. As an aside, it is interesting to note
that, as the circular velocity of the parent galaxy is a few km s$^{-1}$, the
ionized gas (whose sound speed is of order of 10 km s$^{-1}$) will very likely
be able to leave the galaxy, thus quenching further star formation. If 
this is a widespread phenomenon, it might have strong implications for
the evolution of dwarf galaxies as already outlined by some authors  
(Barkana \& Loeb 1999; Ferrara \& Tolstoy 2000). Once the I-front
expands in the IGM, it is channeled -- for similar reasons -- by the large
filaments into the underdense volumes (voids) and its structure becomes
very complex and jagged, assuming, in a
2D representation, a characteristic butterfly shape. 
The typical overdensity of the clumps encountered by the
expanding I-front is $\approx 30$. 
Qualitatively, the typical final extent of the ionized region
is of about 1 comoving Mpc; this size is reached already after 60 Myr.
After that time, the rapid decrease of the source ionization power 
(Fig. \ref{fig2}) slows down the expansion. The
final stage of the evolution is constituted by a relic H$_{\rm II}$ region which
slowly starts to recombine on a time scale which in the voids can be 
as long as 0.8 Gyr.
The MC technique shows here all his power in following the
details of the I-front evolution. For example, it tracks remarkably
well the channeling induced by the large scale structure: the ionization
front cannot propagate inside the densest filaments due to their large
optical depth and short recombination time. In addition, also
the effects of shadowing produced by isolated clumps are clearly recognized 
from the fingers protruding from the H$_{\rm II}$ region to the left of the source
 in the upper panel of Fig. \ref{fig8}.
The ionization cone visible below the source, caused by a large 
overdensity located close to the source in that direction, is also a result of
shadowing. Thus it seems that
our scheme is highly suitable at least for this type of cosmological radiative
transfer calculations. 

We now turn to the differences induced by our two   assumptions
concerning the IMF. Fig. \ref{fig8} shows a comparison between the
final stages (100 Myr after the source turn on) of the I-front evolution
for a Larson   (upper panel) and a Salpeter (bottom) IMF. The source
stellar mass and other properties of the simulations are the same
as those discussed above. For a Salpeter IMF, the volume of the
ionized region is smaller by a factor 8, although the shape is
very similar to the one for a Larson IMF case at an earlier stage,
roughly corresponding to 10 Myr (see Fig. \ref{fig7}). This
was expected from the two adopted SEDs. 
In fact, the total number of ionizing photons (see Fig. \ref{fig2})
per stellar mass formed integrated over the entire 
source lifetime and spectral extent is $5\times 10^{61}$ 
($10^{61}$) for the Larson (Salpeter) IMF.
Thus, the IMF might play an important role  for the reionization of the
universe; in addition, zero-metallicity stars have larger ionizing power
as already stressed previously and recently addressed by other authors
(Cojazzi \etal 2000; Tumlinson \& Shull 2000).  

To assess the differences between the evolution of an I-front
propagating in an inhomogeneous medium, as in the simulations discussed here,
and in an homogeneous gas at the mean IGM density, we have calculated  
the ratio ${\cal V} = V/V_h$ between the ionized volumes in our
simulation and in the homogeneous case. This is found to be roughly
independent of time and equal to ${\cal V}=0.45$ for the Larson IMF.
In the Salpeter IMF case ${\cal V}$ varies slightly with time from 
0.125 to 0.2. Hence, this result confirms that the ionized volume tends to 
be smaller in the inhomogeneous case, 
but suggests that the effect is not dramatic.

\section*{Acknowledgments}

We would like to thank S. Cassisi for providing us with the stellar 
tracks; N. Yoshida and V. Springel for advises in various numerical
problems; the referee T. Abel for useful comments.
This research was supported in part by the National Science 
Foundation under Grant No. PHY94-07194 (SM) and by the Italian
Ministery of University, Scientific Research and Technology (MURST)
(GR).

\label{lastpage}

\newpage


\begin{thebibliography}{99}


\bibitem{} Abel, T., Norman, M. L. \& Madau, P. 1999, ApJ, 523, 66

\bibitem{} Barkana, R. \& Loeb, A. 1999, ApJ, 523, 54 

\bibitem{} Benson, A. J., Nusser, A.,  Sugiyama, S. \& Lacey, C. G. 2000, preprint
(astro-ph/0002457)

\bibitem{} Bianchi, S., Ferrara, A. \& Giovanardi, C. 1996, ApJ, 465, 127

\bibitem{} Bianchi, S., Ferrara, A., Davies, J. \& Alton, P. 2000,
MNRAS, 311, 601

\bibitem{} Brocato, E., Castellani, V., Raimondo, G. \& Romaniello, M.  1999,
A\&AS, 136, 65

\bibitem{} Brocato, E., Castellani, V., Poli F. M. \& Raimondo, G. 2000, A\&A,
submitted

\bibitem{} Bruscoli, M., Ferrara, A., Fabbri, R. \& Ciardi, B.  2000,
MNRAS, 318, 1068                           

\bibitem{} Cashwell, E.D. \& Everett, C.J. 1959,  A Practical Manual on the
Monte Carlo Method for Random Walk Problems (New York: Pergamon)

\bibitem{} Cassisi, S. \& Castellani, V. 1993, ApJSS 88, 509

\bibitem{} Cassisi, S., Castellani, V. \& Tornamb\`e A. 1996, A\&A, 317, 108

\bibitem{} Cen, R. 1992, ApJS, 78, 341

\bibitem{} Chiu, W. A. \& Ostriker, J. P. 2000, preprint (astro-ph/9907220)

\bibitem{} Ciardi, B., Ferrara, A. \& Abel, T. 2000, ApJ, 533, 594 

\bibitem{} Ciardi, B., Ferrara, A., Governato, F. \& Jenkins, A. 2000, MNRAS, 
314, 611, CFGJ

\bibitem{} Cojazzi, P., Bressan, P., Lucchin, F., Pantano, O. \& Chavez,
M. 2000, MNRAS, 315, 51

\bibitem{} Donahue, M. \& Shull, M. J. 1987, ApJ, 232, L13

\bibitem{} Dove, J. B. , Shull, J. M. \&  Ferrara, A. 2000, ApJ, 531, 846

\bibitem{} Ferrara, A., Bianchi, S., Dettmar, R.-J. \& Giovanardi, C. 1996,
ApJL, 467, 69

\bibitem{} Ferrara, A.,  Bianchi, S. Cimatti, A. \& Giovanardi C. 1999,
ApJSS, 123, 437

\bibitem{} Ferrara, A.  \& Tolstoy, E. 2000, MNRAS, 313, 291

\bibitem{} Gnedin, N. Y. 2000, preprint (astro-ph/0002151)

\bibitem{} Gnedin, N. Y. \& Ostriker, J. P. 1997, ApJ, 486, 581

\bibitem{} Haiman, Z. \& Loeb, A. 1998, ApJ, 503, 505

\bibitem{} Haiman, Z., Abel, T. \& Madau, P. 2000, preprint (astro-ph/0009125)

\bibitem{} Kurucz, R. L. 1993, CDRom n. 13

\bibitem{} Larson, R. B. 1998, MNRAS 301, 569

\bibitem{} Miralda-Escud\'e, J., Haehnelt, M. \& Rees, M. R. 2000, ApJ, 530, 1

\bibitem{} Nakamura, F. \& Umemura, M., 1999a, A\&A, 343, 41

\bibitem{} Nakamura, F. \& Umemura, M., in Star Formation 1999b, Proc. of Star
Formation 1999, held in Nagoya, Japan, Ed. T. Nakamoto, Nobeyama Radio
Observatory, p. 28-29

\bibitem{} Norman, M. L., Paschos, P. \& Abel, T. 1998, in H$_2$ in
the Early Universe, Florence, Proc., eds. E. Corbelli, D. Galli \& F.
Palla, p. 455.

\bibitem{} Razoumov, A. \& Scott, D. 1999, MNRAS, 309, 287
 
\bibitem{} Shapiro, P. R. \& Giroux, M. L. 1987, ApJ, 321, L107

\bibitem{} Springel, V., Yoshida, N. \& White, S. D. M. 2000, preprint
(astro-ph/0003162)

\bibitem{} Tumlinson, J. \& Shull, J. M. 2000, ApJ, 528, 65

\bibitem{} Umemura, M., Nakamoto, T. \& Susa, H. 1999, in Numerical Astrophysics, eds. 
Miyama \etal (Kluwer: Dordrecht), p.43

\bibitem{} Valageas, P. \& Silk, J. 1999, A\&A, 347, 1

\bibitem{} Wood, K. \& Loeb, A. 2000, ApJ, in press 


\end{thebibliography}
\end{document}